
\magnification=1200
\baselineskip 18pt

\centerline{\bf Local Density of States and Level Width for Wannier-Stark
Ladders}
\vskip .5in
\centerline {M.C.Chang and Q.Niu }
\vskip .5in
\centerline{Department of Physics, University of Texas, Austin,TX 78712}
\vskip .5in

The local density of states $\rho(x,E)$ is calculated for a Bloch electron in
an electric field.
Depending on the system size, we can see one or more sequences of Wannier-Stark
ladders in
$\rho(x,E)$,  with Lorentz type level widths and apparent spatial localization
of the states.
Our model is a
chain of delta function potential barriers plus a staircase-like electric
potential, with open
boundary condition at both ends of the system.
Using a wave tunneling picture, we find that the level widths shrink to zero as
an inverse power
as the system size approaches infinity, confirming an earlier result.  The
level width will not approach zero if the delta function barriers are replaced
by the
Kronig-Penny potential or smoother ones, as is commonly believed.

\vskip .3in
PACS$:$ 71.20.-b, 71.50.+t, 73.20.Dx, 73.40.Gk\hfil\break

To be published in Phys.~Rev.{\bf B47}, July (or August), 1993.
\vskip 1in
changmc@utpapa.ph.utexas.edu
\vfill\eject
I.~Introduction
\vskip .5cm
It is well known that an electron in an isolated Bloch band will oscillate with
frequency $eFa/h$ in
the Brillouin zone when it is subject to a uniform and constant electric field
$F$, where $a$ is the
lattice constant.  It was first pointed out by Wannier that such  oscillatory
motion is associated
with a ladder like spectrum, now commonly known as Wannier-Stark ladders.
Wannier further proposed
that sequences of such ladders exist even when the  interband coupling is taken
into account.[1]

This proposal was immediately challenged by Zak, who pointed out that sharp
Wannier-Stark ladders cannot exist in a system with any reasonably continuous
periodic potential,
because the spectrum is absolutely continuous.[2]  A numerical study on the
density of states $\rho (E)$ for a system with a cosine potential in an
electric field
also showed  no evidence of the ladder structure [3].  Since the existence
of sharp Wannier-Stark ladders in the one band approximation was rigorously
established [4], it was then suggested that a correct account of the interband
transitions will smear out the discrete spectrum [5].  Further, it has also
been proved that
an isolated group of $n$ Bloch bands can give rise to $n$ sequences of sharp
ladders, so one must
consider an infinite number of coupled bands in order to obtain a continuous
spectrum.
Recently, by decomposing the linear electric potential into
a sawtooth part plus a staircase part, Emin and Hart claimed that the bands may
be
effectively decoupled because the interband matrix
elements of the staircase potential is zero.  This would give a proof for the
existence of a
sharp ladder spectrum, but the arguments leading to the vanishing of interband
matrix elements have
been critcized[6].

On the other hand, experimental and theoretical evidence
in support of the existence of Wannier-Stark ladders also increased during the
same period of time [7]. Krieger and Iafrate clarified the controversies
concerning the use of periodic boundary condition and the neglect of interband
transitions.[8]  Although they did not discuss the energy spectrum at all, they
can
show that the transitions of electrons induced by an incident electromagnetic
field are allowed only if $\Delta E =eFa$, as if these levels exist. In recent
years,
thanks to technological advancement on superlattices, the existence of
a ladder structure seems to be firmly established. [9]

A clear picture about the nature of the W-S ladders should have emerged from
the many years of
contraversial debates: the ladders are not infinitely sharp levels, but they
are resonant
levels whose life times can be long under conditions of small interband
couplings.
In earlier numerical works, such resonant levels are located by searching for
poles of the S matrix or the resolvant in the complex energy plane, with the
level
widths given by the imaginary part of the poles.[7] In this paper,  we reveal
and display the W-S
ladder structure in the local density of states $\rho(x,E)$, which can be
calculated in a simple and
direct way.  As we will show in the main text, the W-S ladders appear naturally
and very clearly in a plot of
$\rho(x,E)$ as sequences of mountains, whose widths in the $E$ direction
represent the level widths, and whose
widths in the $x$ direction represent the localization of the ladder states.
Such structures can be
easily washed out in the total density of states $\rho(E)$.

In order to simplify the calculation, we use in our model a chain of delta
function barriers
to simulate the periodic potential, and use a staircase potential to simulate
the external field.
There has been
criticism concerning the use of the staircase potential, arguing that it will
lead to vanishing
interband transitions [10], however, such criticism is invalid as has been
shown by Kleinman[6] and
others.  Although our model is used mainly for qualitative studies, it can also
be used to simulate
real solids when the spatial extent of the wave functions are large compared to
the lattice constant
as has been demonstarted by Bentosela {\it et al} in a similar situation.[7]

The organization of this paper is as follows: In Sec.2 we describe our model
and derive a formula
for the local density of states. In Sec.3 we present the numerical results for
the local
and total density of states with a variety of parameter values.  In Sec.4, the
behavior of the level
width as a function of system size and other parameters is explained in a
picture of wave tunneling
through the potential barriers.

\vskip .5cm

II.~The Model
\vskip .5cm
We will study the wave function solution of a one dimensional model with
Hamiltonian
$$H=-{\hbar^2\over 2m}{d^2\over dx^2}+V_0a\sum_{j=-N_l}^{N_r}\delta(x-ja)
+V_{ext} \eqno(1) $$
 where $V_{ext}$ is a staircase potential, simulating an external
electric field. Reference of
energy is chosen such that
$$V_{ext}=-j\Delta \ \  {\rm for} \ \ ja<x<(j+1)a\ ,\ j\in Z
\eqno(2) $$
We will take $\hbar^2 \pi^2/2ma^2$ as the unit of energy, therefore
$\Delta =0.2$ corresponds to an electric field of
$F=7\times 10^4\ ({\rm V/cm})$ for $a=200 \AA$ and $m=0.067 m_e.$

We will choose our boundary such that the potential beyond both ends of our
crystal levels off, as shown in Fig.1. The cells are
numbered by the integer $j$ in (2). The wave function in cell $j$ is
$$\psi_j(x)=a_je^{ik_jx}+b_je^{-ik_jx} \eqno(3) $$
where
$k_j=(\pi/a)\sqrt{E+j\Delta}$ (new unit). The coefficients of cell $j+1$ and
cell
$j$ are related by a transfer matrix $\bf{M_j}$ :
$$\left({a_{j+1} \atop
b_{j+1}}\right)={\bf M}_j\left({a_j \atop b_j}\right) \eqno(4) $$
where
$${\bf M}_j={1 \over
2k_{j+1}}\pmatrix{(k_j+k_{j+1}-i\kappa )e^{-i(k_{j+1}-k_j)(j+1)a}
&(-k_j+k_{j+1}-i\kappa )e^{-i(k_{j+1}+k_j)(j+1)a}\cr
(-k_j+k_{j+1}+i\kappa)e^{i(k_{j+1}+k_j)(j+1)a}
&(k_j+k_{j+1}+i\kappa )e^{i(k_{j+1}-k_j)(j+1)a }\cr} \eqno(5) $$
and $\kappa \equiv \pi^2 V_0/a .$ The above formula remains valid
when $E<-j\Delta$, in which case we take $k_j=-i(\pi/a)\sqrt {|j\Delta+E|}.$
Notice that ${\rm det}
{\bf M}_j=k_j/k_{j+1},$ which is not equal to the usual value of unity,
because the velocities corresponding to a given energy in neighboring cells are
different.

In our study we will take $E<N_l\Delta$, and
drop off the exponentially large solution at the left edge by letting
$b_{-N_l-1}=0$. The two solutions $a_{N_r}e^{ik_{N_r}x},b_{N_r}e^{-ik_{N_r}x}$
beyond the right edge correspond to incoming and outgoing waves. The
normalization
of wavefunctions is set by taking $\vert a_{N_r}\vert=1$.
Furthermore, in the actual calculation, the number of cells on the negative $x$
axis is limited to 10
for convenience. The error involved is negligible as long as we are considering
energy only a few
$\Delta$'s large.

The quantity we are after is the local density of states defined by
$$ \eqalign {\rho(x,E)&=\int^\infty_0{{dk \over 2\pi}\mid \psi_k(x) \mid^2
\delta(E-E_k)} \cr &={1 \over {2\pi\hbar}}\sqrt{m\over 2(E+N_r\Delta)}\mid
\psi_k(x)
\mid^2_{E_k=E+N_r\Delta} \cr } \eqno(6) $$
where $E_k=(ka/\pi)^2$
Unlike the total density of states $\rho(E)=\int dx \rho(x,E)$ which is usually
calculated,
the local density of states also contains information about  the location of
states, and is more
appropriate for the purpose of revealing the nature of the W-S ladders.

\vskip .5cm
III.~Result
\vskip .5cm
The coefficients $ a_j, b_j (j=-(N_l+1) \sim N_r) $  are calculated starting
from
$b_{-N_l-1}=0$, and $ a_{-N_l-1} $ is chosen such that $ \vert a_{N_r}\vert=1
$. We can get
$\rho (x,E) $ for some fixed value of energy $E$. The energy is then changed to
a
different value and the calculation is done all over again.
The result is
double checked by calculating $a_j, b_j$'s starting from $a_{N_r} \ {\rm and\ }
b_{N_r}$.

When the electric field is zero, we should see an energy band structure, as
shown in Fig.2.
There are 50 cells ($N_l=10,N_r=40$) in this system. Only the
first band and the lower half of the second band are plotted. The band edges
coincides very well
with a simple theoretical calculation.

Fig.3 is a plot of $\rho (x,E)$ after we turn on the electric field. The
lattice is composed of 50
cells ( labelled $-10 \sim 39 $).  Only 12 cells ($-1 \sim 10$) are plotted
here. The energy range
covered is 3 $\Delta$'s wide. Two sequences of states can be seen. The first
sequence lies at
energies equal to $(0.34+n)\Delta$, where $n=$integer and is equal to 0,1 and 2
in the energy
range of the figure.  The ladder at $E=0.34\Delta$ has a primary peak in cell 1
and 2, with smaller
peaks in the cells to the right.  The
second  sequence is weaker and broader, but is still discernible, and lies at
energies equal to
$(0.65+n)\Delta$ with $n=0,1,2$ in the figure.   The ladder at $E=0.65\Delta$
is primarily peaked in
cell 8, with the heights of the peaks slowly decreasing to the left and right.
It is evident that
$\rho(x,E)$ is invariant under changes $x\to x+a, E\to E-\Delta$ as it should
be.
Each state in both sequences  spans many lattice sites, and will not 'feel'
much
difference whether the electric potential is linear or staircase-like.

The total density of states $\rho (E)$ is obtained by integrating $\rho (x,E)$
with respect to $x$ up to
the right edge of the lattice. The result is shown in Fig.~4 (a) (note the
difference from the result
in the absence of the field shown in Fig.~2). The two peaks within each
$\Delta$ in energy correspond to the two sequences that can be observed in the
local density of
states.  As we enlarge the system, the number of peaks increases (see Fig.~4
(b),(c),(d) ),
corresponding to the increased number of observable sequences.  One can imagine
that as the system
becomes infinite, there will be infinite number of peaks covering up the whole
energy axis, making
the ladder structure unrecognizable in $\rho(E)$.   Fig.~5 is a plot for an
electric field that
is five times larger than in Fig.4, with more peaks showing up in $\rho(E)$
corresponding to more
sequences of observable ladders in the system.

For a system with $(V,\Delta,N_r)=(1,0.2,40)$, two sequences are observed, as
in Fig.3, but with
much narrower level widths. In Fig.6, the spatial extent of the states in these
two sequences are
plotted. One is at $E\simeq 0.3\Delta$, the other is at $E\simeq 0.7\Delta$.
The state in Fig.6(b) is
essentially confined within $3\sim 5$ cells, which is more localized than the
states in the first
'band' in Fig.3.  A cross section of it at $x=3.5a$ is plotted in Fig.7. It is
so sharp that energy
resolution as high as $\Delta /10^{11}$ has to be used. A nice Lorentz curve
appears after such an
immense blow up. The width of the state in Fig.6(a) is about $10^{-6}\Delta$.

Fig.~8 is for another
set up: the lattice has 14 cells ($-10 \sim 3)$, with a
higher barrier $V_0$. It is a cross section of the lowest peak of cell 0. This
peak is essentially
confined within one cell. It is almost as sharp as the peak in Fig.~7(b) even
though we are
considering a relatively small system---there are only 4 cells on the positive
$x$-axis, because the
barriers are much higher. Its width will be calculated by wave tunneling in
Sec.~4.

The level width will decrease as $N$ increases, because it is more difficult
for these resonant states to tunnel out. An interesting question is: when our
system becomes
macroscopically large, will the width shrink to zero or to a finite value? This
is answered in the next section.

\vskip .5cm
IV.~Tunneling and level width
\vskip .5cm
A. General formula

We will be interested in those resonant levels mainly located in a unit cell,
say the lowest level
in cell 0. In the limit of small level width, we may write the level width as
$$\delta E=(\hbar /2) \mid T_0 \mid^2 \nu \eqno(7). $$
where $\nu$ is the attempting frequency in cell 0, and $T_0$ is the
transmission coefficient from
cell 0 to the right edge of the system. It is emphasized that the transmission
coefficient used here
and hereafter refers to the probability currents instead of the probability
amplitudes. The
transmission coefficient through the $j$th barrier is the same for both
directions, and is given by
$$t_j={2i\sqrt{k_{j-1}k_j} \over
i(k_{j-1}+k_j)-\kappa }\ e^{i(k_{j-1}-k_j)aj} \eqno(8) $$
On the other hand, the reflection coefficient of a barrier depends on the
direction of
incidence, and is given by
$$ \eqalign{ r_{j\vdash }&= { i(k_j-k_{j-1})+\kappa \over
i(k_j+k_{j-1})-\kappa }\ e^{-2ik_jaj} \cr r_{j\dashv } &= {
i(k_{j-1}-k_j)+\kappa \over
i(k_{j-1}+k_j)-\kappa }\ e^{2ik_{j-1}aj}\cr} \eqno(9) $$
where $\vdash$ refers to reflection from and to the right, and $\dashv$ from
and to the left.

A general formula for $T_0$ may be obtained through a diagrammatic analysis. We
denote $T_j$ to be
the transmission coefficient from the $j$th cell to the right edge, with the
understanding that all
the paths are contained between the $j$th cell and the right edge. It is then
easy to show that
$$\eqalign {T_0 &=t_1
T_1+t_1\Sigma_1 T_1+t_1\Sigma^2_1 T_1+\dots \cr  &={t_1 \over
1-\Sigma_1}T_1 \cr} \eqno(10)$$
where $\Sigma_1$ is the sum over contributions of \lq self-energy' loops each
of which goes from
cell 1 to the right and resumes the initial position and direction of
propagation only in the final
step. See Fig.~9~(a).

The above formula can be iterated, yielding
$$ T_0 =\prod^{N_r}_{j=1} {t_j \over 1-\Sigma_j} \eqno(11) $$
where $\Sigma_j$ is the total contribution of \lq self-energy' loops starting
and ending in cell $j$.
It is understood that all the paths for $\Sigma_j$ are on the right of cell
$j$, except for the
first and the last steps which are in the $j$th cell.

The formula (11) is an exact result, but the \lq self-energy' contribution
cannot be obtained
exactly. In the case of strong barriers, we may approximate $\Sigma_j$ by
$r_{j\vdash}r_{j+1\dashv}$, coming from a path reflected once by the $(j+1)$th
barrier
and once by the
$j$th barrier. All the other paths must have at least two
transmissions through the $(j+1)$th barrier and are therefore of higher order.
We have done a
calculation for the case of $N_r=4$, $\Delta=0.2$, and $V_0=8$. The half width
is found to be
$6.64\times 10^{-10}$ by the above approximation, while the exact numerical
result as shown in Fig.8
is $(6.4\pm 0.1)\times 10^{-10}$. The error is less than $5\%$.

When the barriers to the right of the $j$th cell are not strong, we cannot use
the above
approximation. However, if the reflection coefficients are small, we only need
to sum over those
loops containing 2 reflections ( Fig.~9~(b) ). Therefore,
$$\Sigma_j\simeq \sigma_j^{(2)}\equiv\sum^{N_r}_{i(>j)}
t_{ji}r_{i\dashv}t_{ij}r_{j\vdash} \eqno(12)$$
where $t_{ji}=t_{j+1}\cdots t_{i-1}$ for $j<i$, $t_{1j}=t_1\cdots t_j$,
$t_{jN}=t_{j+1}\cdots t_N$.
$$T\simeq T'=\prod^{N_r}_{j=1} {t_j \over 1-\sigma_j^{(2)}} \eqno(13)$$
It is interesting to note that the above formula is even valid in the
opposite limit of small $t$'s, because then only the $i=j+1$ term survives.
This result is
identical to the approximate formula that we derived before for the strong
barrier case.
\vskip .5cm

B. $\delta-$function barriers.

We are now ready to calculate the energy level width of a $\delta$ potential
chain subjected
to electric field. When $k_j\gg\kappa$, which is true for a large $j$ such that
$j\Delta +E\gg
V_0^2$,  $$ \vert t_j \vert \simeq { 1 \over \sqrt{ 1+(\pi
V_0/2)^2(1/(E+j\Delta))}} \eqno(14) $$
$$\vert r_{j \vdash} \vert \simeq \vert r_{j\dashv} \vert \simeq {\pi V_0 \over
2}{1
\over \sqrt{E+j\Delta} } \eqno(15) $$
If we are considering the limit $\Delta \rightarrow 0$, (14), (15) remain valid
if $E\gg V_0^2$.

Using these approximations, we will find that : (phase factors are included)
$$ \sigma_j^{(2)}
\simeq -{\alpha \over \Delta}(j+E/\Delta)^{\beta-1}\left( \sum^{N_r}_{l=j}
{1 \over (l+E/\Delta)^\beta}\ e^{i\xi {l^{3/2}}}\right) e^{-i\xi {j^{3/2}}}
\eqno(16) $$
where  $ \xi \equiv (4\pi /3) \sqrt{\Delta},\ \beta\equiv
1/2+\alpha/\Delta,\ \alpha \equiv (\pi V_0/2)^2 $. This expression is correct
as long as $N\Delta$
is larger than, or of the same order of magnitude as $E$.

The problem now is how to sum up this series? If $\Delta$ is not too small, the
difference
$\delta(\xi l^{3/2})\simeq (3/2)\xi l^{1/2}=2\pi\sqrt{l\Delta}\gg 2\pi$. In
this case $\xi
l^{3/2}$ mod $2 \pi $ can be considered random. Therefore, ( $\sigma_j^{(2)}$
is simply written as
$\sigma_j$, and $N_r$ is written as $N$ in the following equations )
$$ \vert \sigma_j \vert \simeq \sqrt{\langle {\vert \sigma_j
\vert}^2 \rangle } \simeq \sqrt{\alpha \over 2\Delta } (j+E/\Delta)^{-1/2}
\eqno(17) $$
where $\langle \ \rangle$ means taking the average with respect to the angle
of the complex number. We assumed $N$ is so large that the term $(N+{E \over
\Delta})^{-1/2}$ is
negligible.

The quantity $\vert \sigma_j \vert$ is small since we are considering $j > J$.
The
infinite product $\Pi (1-\sigma_j)^{-1}$ in (13) can then
be approximated by $ e^{{\sum \atop j}{\sigma_j\atop\ }} $ when the index $j$
is larger
than $J$. The phase of $\sigma_j$ is , again, a random number. Therefore, we
get
$$ \vert \sum_{j>J}^N \ \sigma_j \vert \simeq \sqrt{ \left\langle {\vert
\sum_{j>J}^N \  \sigma_j \vert}^2 \right\rangle }\simeq \sqrt{\alpha \over
2\Delta} \left( {\ln}({N\Delta+E \over J\Delta+E}) \right)^{1/2} \eqno(18) $$
with a random phase factor $e^{i\theta}$.

Finally, we get
$$\eqalign{ \vert T'\vert
&= C_J \left\vert\prod_{j>J}^N
{t_j} \prod_{j>J}^N { 1 \over 1-\sigma_j}\right\vert \cr
&\simeq C_J\exp\left( \sum_{j>J}^N \ln \vert {t_j} \vert \right)
\left\vert\exp\left( \sum_{j>J}^N \sigma_j \right) \right\vert \cr
&\simeq C_J\exp\left( -{\alpha  \over 2\Delta} \ln ({N\Delta+E \over
J\Delta+E}) +\sqrt{{\alpha \over 2\Delta}\ln({N\Delta+E \over J\Delta+E})} \cos
(\theta_{rnd}) \right) \cr}\eqno(19)  $$
in which $C_J$ is a finite number.
We can see that $T'$ will approach zero as $ N \rightarrow
\infty $

$T'$ is a good approximation to correct $T$, at least up to the first order of
approximation. It is improbable that higher order corrections will lead to a
nonzero $T$ since that
would require the correction of the part of $ \prod{ (1-\sigma_j)^{-1}} $
diverges at
least as fast as ${\exp}((\alpha/ 2\Delta) {\ln}(N/J)) $.
Notice that the dominant term $-(\alpha / 2\Delta)\ln
\left({N\Delta+E \over J\Delta+E} \right)$ comes solely from the zeroth
order contribution. The wavefunction $\psi(x)\sim N^{-{\alpha \over 2\Delta}}$
is power law
localized, which is a marginal case between extended and exponentially
localized cases.
Similar studies of localization in the energy space also showed marginal
behavior [11].

\vskip .5cm
C.~The Kronig-Penny type barriers

The same principle can be applied to the Kronig-Penny type barriers.
Assume that the width and height of the
barriers are $a_1$ and $V_0$ respectively, that the distance between barriers
is
$a_2$, that $a=a_1+a_2$ is the size of unit cell, and that the potential drops
one $\Delta$ over
distance $a$. Then, a straightforward calculation shows that, when $j$ is much
larger
than one
$$\vert t_j \vert \simeq \left\vert 4 k'_j/k_j \over
(1+k'_j/k_j)^2-(1-k_j'/k_j)^2
e^{2ik_ja_1} \right\vert. \eqno(20) $$
where $ k'_j=\sqrt{2m(E+j\Delta)}/\hbar$; $k_j=\sqrt{2m(E+j\Delta-V_0})/\hbar$.

When $E+j\Delta\gg V_0$,
$$ {k'_j \over k_j} \simeq 1+{1 \over 2}{V_0 \over E+j\Delta}
\eqno(21)$$
Equation (20) can be further simplified :
$$ \vert t_j \vert \simeq 1- \left({V_0 \over
4(E+j\Delta)}\right)^2(1-\cos(2k_ja_1))
\eqno(22)$$
Therefore,
$$ \prod_{j=1}^\infty \vert t_j \vert \simeq \left(\prod^{J-1}_{j=1} \vert t_j
\vert\right) \exp
\left\{-({V_0 \over4})^2 \sum_{j=J}^\infty  {1 \over (j\Delta+E)^2
}(1-\cos(2k_ja_1))\right\}
\eqno(23)$$
The exponent is a finite quantity. Therefore, up to zeroth order,
the level width is finite as $N \rightarrow \infty.$ This is true as long as
the
height of barriers is finite.

The summation in the exponent of (23) can be approximated by an integral.
It turns out that,
$$\prod^\infty_{j=1} \vert t_j \vert \sim \exp \left\{ -\alpha\left({V_0 \over
4}
\right)^2 {1 \over \Delta(J\Delta+E)} \right\} \eqno(24)$$
where $\alpha$ is a number of order 1.
We can see that, as $\Delta$ becomes very small ( but not smaller than $E/N$ ),
the level width
shrinks to zero with a rate $\sim \exp(-1/\Delta)$. In the $\Delta \rightarrow
0$ limit, (24) is no
longer valid.  In this case, $j\Delta+E \simeq E$, and the tunneling
coefficient turns out to be
$\simeq \exp(-\alpha(V_0/4)^2(N/E^2))$,  which is finite as long as the system
is finite.

We can also reproduce the leading term in the exponent of (19) for the case of
the Kronig-Penny model
in the limit  that $a_1 \rightarrow 0,\ V_0 \rightarrow \infty$, and at the
same time keep
$a_1V_0$ a constant.

\vskip .5cm
V. Conclusion
\vskip .5cm

We have calculated the local density of states $\rho(x,E)$ for a 1D model using
delta function
potential barriers to simulate a periodic potential and a staircase potential
to simulate an external
electric field.   The W-S ladders are revealed most clearly as peaks of the
local density of states in
3D plots of $\rho(x,E)$.  We have demonstrated that the ladder structures can
be easily washed out in
the total density of states.

We have also offered a new approach, based on the calculation of leakage
current
in a graded array of potential barriers, to evaluate the level width.[12] This
complements
the usual method based on the picture of Zener tunneling between Bloch
bands.[13]  It is
shown that, when the system size $N$ is very large, the level width $\delta E
\sim N^{(-\alpha /
\Delta)} $ in the case of $\delta $ barriers. We choose $\delta$ potential
barriers
to simplify the  derivation of the transfer matrices as well as transmission
and reflection
coefficients.  The same approach can be applied to nonsingular potentials.
We found that, when the potential barriers are  rectangular, the reflection
coefficient
vanishes like $1/x$ for large $x$, and that the energy level width will not
shrink to zero as the
size of the system becomes infinite.  When the potential is smoother, the
reflection  coefficient
vanishes even faster.  The electron  wave function can be localized
only in the case of a $\delta$ potential chain, but even then the localization
is found to be
only marginal; that is, it obeys a power law.

Some of the general properties of W-S ladders are summarized below, and these
should also be valid in
a real solid. Firstly, the number of sequencies presented in a finite system
can be estimated from a
tilted-band diagram. It can be controlled either by changing the size of the
system, or, more
conveniently, by changing the electric field.  Secondly, among these sequences,
there is one that
is most pronounced.  This is best illustrated in Fig.~3~(d). The states of the
sequence that are closest to the staircase are sharper and more localized
spatially. This can be
explained by the theory in Sec.4. The states in this sequence have initial
barriers which are
difficult to tunnel through. The states in the other sequences are more remote
from the origin,
and therefore have higher energy with respect to the ground. This makes them
easier to tunnel
through. Thirdly, within the same sequence, the states in cell --1 are narrower
than those in cell 0,
and those in cell 0 are narrower than those in cell 1...etc. This is because
the state in cell --1
has to go through more barriers in order to tunnel out. A similar kind of
asymmetry leads to the
asymmetrical absorption spectra in the experiment that was done by
Agull\'o-Rueda {\it et al} [9].

\vskip .5cm
{\bf Acknowledgment}
\vskip .5cm
The authors wish to thank W.Kohn, L.Kleinman and P.Ao for many valuable
discussions.
This work is supported in part by the Texas Advanced Research Program, by a
Precision
 Measurement grant from NIST, and by a fellowship on the Trull Centennial
Professorship
at the University of Texas at Austin.

\vskip .5cm
{\bf References}
\vskip .5cm
\item {1.} G.H.Wannier, {\it Phys. Rev.} {\bf 117}, 432 (1960).
\item {2.}J.Zak, {\it Phys. Rev. Lett.} {\bf 20}, 1477 (1968)
; A.Rabinovitch {\it Phys. Lett.} {\bf 33A}, 403 (1970) ; J.E.Avoron, J.Zak,
A.Grossmann and L.Gunther, {\it
J.Math.Phys.}  {\bf 18}, 918 (1977).
\item {3.}A.Rabinovitch and J,Zak, {\it Phys. Rev.} {\bf B4}, 2358 (1971).
\item {4.}K.Hacker and G.Obermair {\it Z. phys.} {\bf 234}, 1 (1970).
\item {5.}A.Rabinovitch and J.Zak {\it Phys. Lett.} {\bf 40A},189 (1972) ;
J.N.Churchill and
F.E.Holmstrom, {\it Phys. Lett.} {\bf 85A} 453 (1981).
\item {6.}D.Emin and C.F.Hart, {\it Phys. Rev.} {\bf B36}, 7353
(1987);  L.Kleiman, {\it Phys. Rev.} {\bf B41}, 3857 (1990)
; P.N.Argyres and S.Sfiat, {\it Phys. Lett.} {\bf 146A}, 231 (1990)
; J.Zak, {\it Phys. Rev.} {\bf B43}, 4519 (1991).
\item {7.}S.Maekawa, {\it Phys. Rev. Lett.}{\bf 24}, 1175 (1970);  R.Koss and
L.Lambert, {\it Phys.
Rev.} {\bf B5}, 1479 (1972). J.R.Banavas and D.D.Coon, {\it Phys. Rev.} {\bf
B17}, 3744 (1978) ;
F.Bentosela, V.Grecchi and F.Zironi, {\it J.Phys. C: Solid State Phys.} {\bf
15}, 7119 (1982).
\item {8.}J.B.Krieger and G.J.Iafrate, {\it
Phys. Rev.} {\bf B33}, 5494 (1986).
\item {9.}To quote a few of them:
 J.Bleuse, G.Bastard and P.Voison, {\it Phys.
Rev. Lett.} {\bf 60}, 220 (1988) ;
F.Agull\'o-Rueda, E.E.Mendez and J.M.Hong, {\it Phys. Rev.} {\bf B40}, 1357
(1989) on optical absorption ; explained via exiton effect by M.M.Dignam and
J.E.Sipe,
{\it Phys. Rev. Lett.} {\bf 64}, 1797 (1990). M.Nakayama, {\it Phys. Rev.
Lett.} {\bf 65},
2720 (1990) on electroreflectance.
M.K.Saker, {\it Phys. Rev.} {\bf B43}, 4945 (1991) on
photoconductivity spectra.
\item {10.}J.Zak, {\it Phys. Lett.} {\bf 76A}, 287 (1980).
\item {11.}G.Blatter and D.A. Browne, {\it Phys. Rev.} {\bf B37}, 3856 (1988) ;
P.Ao {\it Phys. Rev} {\bf B41}, 3998 (1990).
\item {12.}It is interesting to observe that similar phenomenon as W-S ladders
can also be
found if electron wave function is replaced by EM wave, see G.Monsivais and
M.Castillo-Mussot,
{\it Phys. Rev. Lett.} {\bf 64}, 1433 (1990).
\item {13.}A.M.Berezhkovskii and A.A.Ovchinnikov, {\it Sov. Phys. Solid State},
{\bf 18}, 1908 (1976).

\vfill\eject

\centerline{\bf Figure Captions}
\vskip .5cm
Fig.1. The potential of the system described in equations (1),(2).
It corresponds to the case $N_r=N_l=5$.

Fig.2. The plot of $\rho (E)$ when electric field is zero,
$(V_0,\Delta,N_r)=(0.2,0,40)$.
$N_l$ in this and in all of the following figures are fixed to be 10.

Fig.3~(a),(b). Local density of states $\rho (x,E)$ plotted as a function of
$x$ and $E$.
$(V_0,\Delta,N_r)=(0.2,0.2,40)$ 12 cells are plotted (--1$\sim$10). Range of
energy
covered is 3 $\Delta$, spatial resolution is $a/5$, energy resolution is
$\Delta/100$.
The curves will be smoother with higher resolution.

Fig.4~(a),(b),(c),(d). A series of plots of $\rho (E)$ for a system with
$(V_0,\Delta)=(0.2,0.2)$,
$N_r=40,60,100,160$ respectively. (a) is for the system plotted in Fig.~3. Also
note the difference
between (a) and Fig.~2.

Fig.5~(a),(b). Two plots of $\rho (E)$ for a system with
$(V_0,\Delta)=(0.2,1)$,
$N_r=40,160$ respectively. The electric field of the systems in this figure is
5 times larger
than those in Fig.~4.

Fig.6. Plots of the states at $E_1=0.339063\Delta$, Fig.(a) ; and
$E_2=0.6757468519\Delta$, Fig.(b),
in a system with $(V,\Delta,N_r)=(1,0.2,40)$.

Fig.7. A cross section of the peak in cell 3 ($x=3.5a$) in Fig.~6(b). The range
of energy covered
is extremely small. $E_c=0.67574685190 \Delta$, half width $\delta E=6\times
10^{-11}\Delta$.

Fig.8.  A cross section of the lowest peak in cell 0.
$(V_0,\Delta,N_r)=(8,0.2,4)$. $E_c=4.7607583641\Delta$, $\delta E=6.4\times
10^{-10}\Delta$. Notice
that the strength of the $\delta$ function barriers is 8 time large. The number
of cells on the
positive $x$ axis is only 4. The level lies only slightly below the level of
an infinitely deep
square well, which is 5 $\Delta$.

Fig.9. (a) A path that makes its last return at point $j$. We call the
bold-faced part
a self energy loop of $j$. $\Sigma_j$ is the summation of every possible self
energy loops of $j$.
(b) A self energy loop with only two reflections. The associated tunneling
amplitude
is stated in (12). $\sigma_j^{(2)}$ is the summation of every possible self
energy loops
of this kind.
\vfill\eject
\end